\documentclass[twocolumn,showpacs,preprintnumbers,amsmath,amssymb,superscriptaddress]{revtex4-1}
\usepackage{graphicx}% Include figure files
\usepackage{dcolumn}% Align table columns on decimal point
\usepackage{bm}% bold math
\renewcommand{\vec}[1]{\boldsymbol{#1} }

\begin{document}

\title{Dynamics of wet granular hexagons}

\author{Manuel Baur}
\author{Kai Huang}
\email{kai.huang@uni-bayreuth.de}
\affiliation{Experimentalphysik V, Universit\"at Bayreuth, 95440 Bayreuth, Germany}

\date{\today}

\begin{abstract}

The collective behavior of vibrated hexagonal disks confined in a monolayer is investigated experimentally. Due to the broken circular symmetry, hexagons prefer to rotate upon sufficiently strong driving. Due to the formation of liquid bridges, short-ranged cohesive interactions are introduced upon wetting. Consequently, a nonequilibrium stationary state with the rotating disks self-organized in a hexagonal structure arises. The bond length of the hexagonal structure is slightly smaller than the circumdiameter of a hexagon, indicating geometric frustration. This investigation provides an example where the collective behavior of granular matter is tuned by the shape of individual particles.

\end{abstract}

\maketitle

%%%%%%%%%%%%%%%%%%%%%%%%%%%%%%%%%%%%%%%%%%%%%%%%%%%%%%%%%%%%
%                  Introduction
%%%%%%%%%%%%%%%%%%%%%%%%%%%%%%%%%%%%%%%%%%%%%%%%%%%%%%%%%%%%

Each grain of sand has a unique shape~\cite{Greenberg2008}. Understanding how shape matters in the collective behavior of granular matter~\cite{Baule2013,Neudecker2013, Boerzsoenyi2013,Athanassiadis2013} is crucial for geophysical and industrial applications~\cite{Acevedo2014,Szabo2015,Parteli2016}. For instance, a change from spherical to ellipsoidal shape effectively enhances the packing density~\cite{Donev2006}. Due to dissipative particle-particle interactions~\cite{Duran2000,Brilliantov2004a,Gollwitzer2012,Mueller2016}, continuous energy injection is necessary to keep granular matter in various nonequilibrium stationary states (NESS) that share common features with their equilibrium counterparts~\cite{Jaeger1996,Huang2009b,May2013,Huang2015}. Such features indicate the possibility of extending recent advances on shape mediated self-assembly of thermally driven particles~\cite{Damasceno2012, Gantapara2013, Spellings2015} to athermal systems such as granular matter. 

In a granular monolayer, understanding the collective dynamics of spherical particles still remains a challenge~\cite{Olafsen2005,Clerc2008,Castillo2012,Komatsu2015}. For elongated particles, analogues to liquid-crystal (LC) mesophases~\cite{Galanis2006,Narayan2006,Mueller2015,Walsh2016}, collective swirling~\cite{Aranson2007}, and giant number fluctuations~\cite{Narayan2007} have been investigated extensively. Self-propelled particles with polar asymmetry have been used to understand the collective dynamics of active matter~\cite{Kudrolli2008,Deseigne2010}. Following recent advances on thermally driven platelets with polygonal shapes, which yield interesting LC and rotator-crystal (RC) mesophases~\cite{Zhao2009,Zhao2011,Zhao2012,Nguyen2014,Zhao2015}, it is intuitive to explore the nonequilibrium counterparts for identifying the universal and non-universal aspects in the collective behavior of anisotropic particles.

Here, we show that hexagonal disks confined in quasi-two-dimensions prefer to rotate upon excitation and the rotators self-organize into a hexagonal crystal upon wetting, while the translational order is weakly dependent on driving. The geometric frustration induced by cohesion leads to cooperative rotations and non-trivial collective behavior. The preference to rotate arises from the broken circular symmetry of the hexagonal shape and the rotation speed can be predicted analytically, suggesting the possibility to controllably excite the rotational degrees of freedom of particles via shape design.

%%%%%%%%%%%%%%%%%%%%%%%%%%%%%%%%%%%%%%%%%%%%%%%%%%%%%%%%%%%%
%                  Methods
%%%%%%%%%%%%%%%%%%%%%%%%%%%%%%%%%%%%%%%%%%%%%%%%%%%%%%%%%%%%

\begin{figure}[b]
\includegraphics[width = 0.45\textwidth]{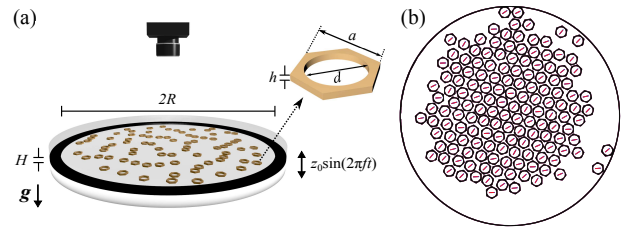}
\caption{\label{fig:setup}(color online) (a) A sketch of the experimental setup with definitions of disk dimensions. The container is vibrated sinusoidally against gravity. (b) A representative image of positionally ordered disks with their orientations marked with red (gray) bars captured with $f=50$\,Hz, $\Gamma=30$, and $W=1\%$.}
\end{figure}
 
Figure~\ref{fig:setup}(a) is a sketch of the experimental setup. The particles are cut from a brass (CuZn39Pb5, density $\rho=8.5\,{\rm g}\cdot{\rm cm}^{-3}$) hexagonal rod with a cylindrical hole of diameter $d=8$\,mm in the center. They have the same inscribed circle diameter $a=10$\,mm and height $h=2$\,mm. After mixed with $V_{\rm liq}$ of purified water~(\mbox{LaborStar TWF}, surface tension $\gamma=0.072$\,N/m), $N=150$ particles are filled in a cylindrical polycarbonate container with height $H=1.5h$ to ensure a monolayer. The inner radius is $R=9.5$\,cm, corresponding to a global area fraction $\phi\approx 46\%$. The liquid content is defined as $W=V_{\rm liq}/(NA_{\rm p}h)$ with $A_{\rm p}=3\sqrt{3}a^2/2-\pi d^2/4$ the base area of a disk. The bottom and lid of the container are $2$\,cm thick to avoid bending of the container upon vibrations. The container is driven sinusoidally by an electromagnetic shaker (Tira TV50350) with frequency $f$ and amplitude $z_{0}$ controlled with a function generator (Agilent FG33220). The dimensionless acceleration $\Gamma=4\pi^2 f^2 z_{0}/g$, with $g$ the gravitational acceleration, is measured with an accelerometer (Dytran 3035B2). In order to distribute the wetting liquid homogeneously and to minimize the memory effect from particles sticking on the container due to cohesion, we apply a high $\Gamma=50$ at $f=75$\,Hz before each experimental run. Using back-light LED illumination, high-contrast images of the particles are captured with a synchronized high-speed camera (IDT MotionScope M3). The captured images are subjected to an analysis algorithm that detects the position and orientation of each particle. 

%%%%%%%%%%%%%%%%%%%%%%%%%%%%%%%%%%%%%%%%%%%%%%%%%%%%%%%%%%%%
%                  Results and Discussions
%%%%%%%%%%%%%%%%%%%%%%%%%%%%%%%%%%%%%%%%%%%%%%%%%%%%%%%%%%%%

\begin{figure}
\includegraphics[width = 0.45\textwidth]{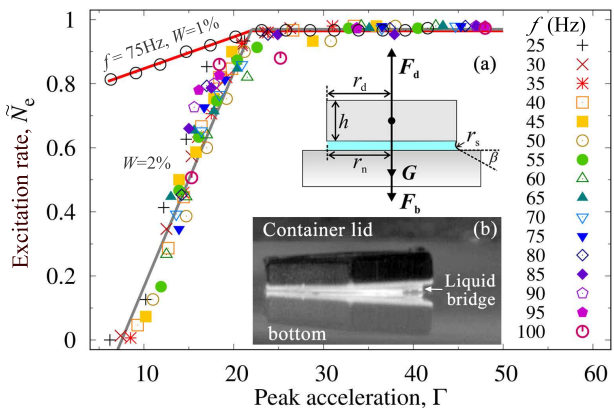}
\caption{\label{fig:nmob}(color online) The excitation rate $\tilde N_{\rm e}$ as a function of $\Gamma$ at different $f$ and $W$. The solid lines are fits to the data at different $W$. Inset (a): Sketch of a liquid bridge formed between a disk and a horizontal plane. (b): Side view image of a hexagonal disk detaching from the container bottom, which separates the disk from its mirror image.}  
\end{figure}

Because of the capillary force $\vec F_{\rm b}$ induced by the liquid bridge (insets of Fig.~\ref{fig:nmob}), sufficiently high $\Gamma$ is necessary to excite the disks (i.e., detach from the container and rotate spontaneously)~\cite{suppl1}. As shown in Fig.~\ref{fig:nmob}, the excitation rate $\tilde N_{\rm e}=N_{\rm e}/N$ with $N_{\rm e}$ the number of immobile particles grows with $\Gamma$ until it saturates at a critical acceleration $\Gamma_{\rm c}$. The collapse of data for all $f$ at $W=2\%$ suggests that $\Gamma$, which determines the force acting on the particles, dominates the excitation process. In order to quantify the threshold, we fit the data with $\tilde N_{\rm e}(\Gamma)=a(\Gamma-\Gamma_{\rm c})+b$ if $\Gamma \le \Gamma_{\rm c}$, and $b$ otherwise, which yields $a=0.067\pm0.006$, $b=0.970\pm0.015$ and $\Gamma_{\rm c}=22\pm4$. It shows that $\approx97\%$ of the hexagonal disks are excited above the threshold $\Gamma_{\rm c}\approx22$. Reducing the liquid content to $W=1\%$ leads to the same threshold $\Gamma_{\rm c}$, but a smaller slope $0.010\pm0.001$.

Quantitatively, $\Gamma_{\rm c}$ can be understood from the force balance $\vec F_{\rm d}=\vec G + \vec F_{\rm b}$, where $\vec G$, $\vec F_{\rm d}=\Gamma \vec G$, and $\vec F_{\rm b}$ are gravity, maximum driving force, and the capillary force, respectively. For a cylinder with radius $r_{\rm d}$ sticking on a wet plane, the normal capillary force can be estimated with $F_{\rm b}=2\pi\gamma r_{\rm n} \cos\beta + \Delta p_{\rm L}\pi r_{\rm n}^2$ with contact angle $\beta$, neck radius of the liquid bridge $r_{\rm n}$ and Laplace pressure $\Delta p_{\rm L}\approx \gamma(r_{\rm s}^{-1}-r_{\rm n}^{-1})$~\cite{Butt2009}, where $r_{\rm s}$ corresponds to the curvature along the meridional bridge profile. Because the lower limit of the separation distance is the surface roughness $\epsilon$, we estimate $r_{\rm s}=\epsilon/\cos\beta$, which yields $\approx20\,\mu$m for the experimental condition with $\beta\approx 65^{\circ}$~\cite{Gajewski2008}. Note that this is only a rough estimation because in reality neither $\epsilon$ nor $\beta$ is a constant due to abrasion and contact angle hysteresis~\cite{Herminghaus2006}. Because $r_{\rm s} \ll r_{\rm n}$, the above estimation can be simplified into $F_{\rm b}=\gamma A/r_{\rm s}$ with $A$ the neck area of the liquid bridge. Note that the shape of the disk does not play a dominating role in this simplified form. Consequently, the critical acceleration is

\begin{equation}
\label{eq:critGam}
\Gamma_{\rm c}=\frac{F_{\rm b}}{G}+1 = \frac{\gamma}{\rho r_{\rm s} h g}\frac{A}{A_{\rm p}}+1.
\end{equation}    

\noindent Because of imperfect wetting, the factor $A/A_{\rm p}$ varies from particle to particle, leading to a range of critical acceleration to excite the disks. An estimation of the maximum critical acceleration with $A/A_{\rm p}=1$ yields $22.6$, which agrees with $\Gamma_{\rm c}$ obtained from the experiments. Eq.~\ref{eq:critGam} also indicates the independence of $\Gamma_{\rm c}$ on $W$. For $\Gamma<\Gamma_{\rm c}$, larger $\tilde N_{\rm c}$ (i.e., smaller slope $a$) is expected for $W=1\%$ than $W=2\%$ due to less contact area $A$ covered by the liquid. 

\begin{figure}
\includegraphics[width = 0.42\textwidth]{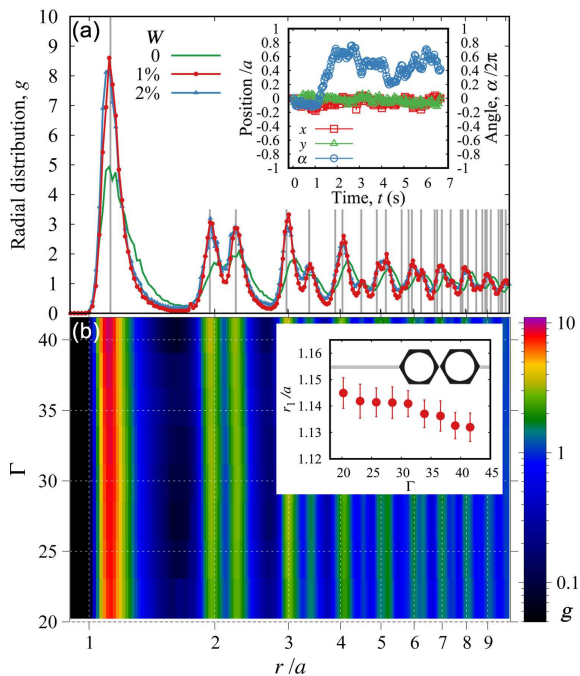}
\caption{\label{fig:gr}(color online) (a) Radial distribution function, $g(r/a)$ of $N=150$ particles driven with $f=75$\,Hz and $\Gamma=25$ for both dry and wet cases. The gray bars mark all possible disk-disk distances for a perfectly hexagonal structure. The logarithmic scale of $r/a$ is chosen to highlight the deviation of the first peak from $r/a=1$. The inset shows the position and orientation of a representative disk with time. (b) $g(r/a)$ for all $\Gamma>\Gamma_{\rm c}$ at fixed $f=75$\,Hz and $W=1\%$. Inset shows the rescaled mean neighboring distance $r_{\rm 1}/a$ as a function of $\Gamma$. The gray line marks the minimum distance of two freely rotating disks.}
\end{figure}

After excitation, the disks self-organize into an ordered state~\cite{suppl1} [see Fig.~\ref{fig:setup}(b)], which is robust in the sense that different initial configurations yield the same structure~\cite{suppl2}. As shown in Fig.~\ref{fig:gr}(a), the positional order of the excited disks is characterized with the radial distribution function $g(r/a)$ with $r/a=1$ the edge-edge contact distance. $g(r/a)$ is obtained through an average over all frames captured in the stationary state. For the wet case, the first peak location is $\approx 1.141$, slightly smaller than the circumdiameter of a disk $2a/\sqrt{3}\approx 1.155a$ (i.e., the minimum distance between two freely rotating disks), indicating a slight hindrance of the rotation by the neighbors. The disks change their sense of rotation through intermittent interactions with their neighbors or the container~\cite{suppl1}, leading to a stepwise change and fluctuations of the angle [inset of Fig.~\ref{fig:gr}(a)].
This geometric constraint is named `frustration' in thermally driven colloidal systems~\cite{Zhao2009,Zhao2011}. Note that this constraint cannot be attributed to the tilting of the disks because the limited tilting angle $\theta \le 5.8^{\circ}$ leads to a length contraction $\le0.003a$. A comparison to the perfectly hexagonal structure with the same bond length [gray bars in Fig.~\ref{fig:gr}(a)] clearly illustrates the crystalline structure formed. Moreover, there is an overlap between $g(r/a)$ obtained with different $W$, indicating that the positional order is weakly dependent on the amount of wetting liquid added. 

A close view of the liquid dynamics indicates that the capillary bridges only form temporarily due to rotation, i.e., there are no permanent cohesive interactions between neighboring particles, different from a wet granular crystal composed of spheres~\cite{May2013}. Consequently, the influence of $W$ on the translational order is weakened. Without the wetting liquid, the rotating disks may form clusters due to frequent inelastic collisions as they approach each other, leading to the agglomeration of particles~\cite{Goldhirsch1993}. However, the translational order is much less pronounced than the wet cases.

Surprisingly, as shown in Fig.~\ref{fig:gr}(b), the control parameter $\Gamma$ does not influence the ordered structure. Moreover, the ordered state also persists as $f$ varies from $50$\,Hz to $100$\,Hz, indicating that the detailed balance between energy injection and dissipation does not play an essential role in determining the NESS. Instead, as will be discussed below, the particle shape matters. The average distance between neighboring particles $r_{\rm 1}/a<1.155$ [inset of Fig.~\ref{fig:gr}(b)] again suggests geometric frustration. A comparison to the dry case, which yields $r_{\rm 1}/a \approx 1.157 > 1.155$, indicates that the geometric frustration arises from cohesion. As $\Gamma$ grows, the slight decrease of $r_{\rm 1}/a$ suggests that the disks tend to order more closely. This counter-intuitive behavior can be understood from the reduced $\theta$ at high $\Gamma$, which leads to a higher chance of forming liquid bridges and thus a stronger influence of cohesion. As more liquid is added, the compaction is more pronounced, until eventually the rotational degrees of freedom are restricted and a transition into a crystalline state arises. A more quantitative characterization of transitions between various NESS will be a focus of further investigations.

\begin{figure}
\centering
\includegraphics[width = 0.4\textwidth]{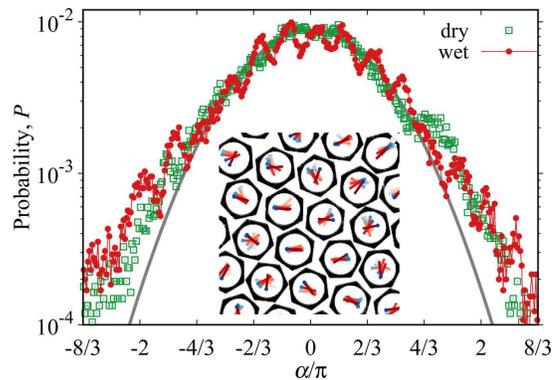}
\caption{\label{fig:ang}(color online) Probability of angle distance, $P(\alpha)$ for dry and wet ($W=1\%$) disks driven at $\Gamma=45$ and $42$, respectively. The gray line is a fit to the data for dry disks with a normal distribution $P_{\rm 0}(\alpha)=p_{\rm 0}e^{-(\alpha/\sigma_{\rm 0})^2}$, which yields $p_{\rm 0}=0.010$ and $\sigma_{\rm 0}=2.73$. Inset is a snapshot of the disks in the ordered state with the evolution of their positions (center of the bars) and orientations in the past $40$ vibration cycles marked with red(gray) bars with different intensities. Lighter colors correspond to earlier time. Other parameters are the same as in Fig.~\ref{fig:gr}.}
\end{figure}

Figure~\ref{fig:ang} shows the influence of geometric frustration on the rotational degrees of freedom of rotating disks. As indicated by a close view of the dynamics in the ordered state (inset of Fig.~\ref{fig:ang}), the disks rotate while being caged in the hexagonal structure. The rotation of the disks can be either hindered or not, depending on the interactions with their neighbors. The hindered rotational degrees of freedom are clearly illustrated with the distribution of the angular distance $P(\alpha)$ for all excited disks with respect to the direction of the hexagonal crystal formed, which fluctuates slightly within $1.5$ degrees over $500$ vibration cycles. For both dry and wet cases, $P(\alpha)$ can be fitted with a normal distribution in the range $\alpha\in[-4\pi/3,4\pi/3]$, in agreement with the prediction of the central limit theorem considering the disks as random walkers in the angular direction. However, there exists a prominent deviation from the fit at larger $|\alpha|$, demonstrating the non-trivial aspect in the nonequilibrium system. Analyzing the distribution in connection to other nonequilibrium systems with `rotors'~\cite{Altshuler2013,Snezhko2015,Scholz2016} will be a focus of further investigations. For the wet case, the pronounced modulation of $P(\alpha)$ with respect to the fit shows the tendency for the disks to align at multiples of $\pi/3$. This feature again suggests the geometric frustration induced by cohesion. A comparison over different $\Gamma$ shows that larger $\Gamma$ leads to stronger geometric frustration, in agreement with the above analysis of $r_{\rm 1}/a$.  

\begin{figure}
\includegraphics[width = 0.4\textwidth]{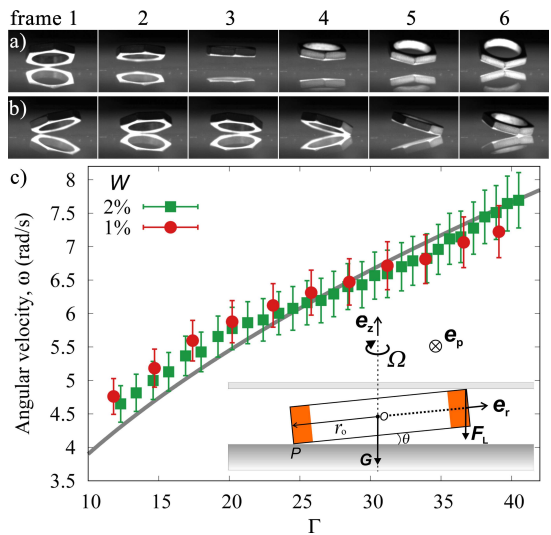}
\caption{\label{fig:rot}(color online) (a) and (b) are series of images showing the clattering and precession of a hexagonal disk on a dry vibrating plane ($f=50$\,Hz, $\Gamma=1.20$) with a time step of $4$\,ms. (c) Mean angular velocity as a function of $\Gamma$ for both liquid contents. The gray line corresponds to the prediction of Eq.~\ref{eq:omega}. The error bars correspond to the standard error. Inset: A disk precesses about the vertical axis $\vec e_{\rm z}$ with an angular velocity $\vec \Omega$. }
\end{figure}

To understand why the disks prefer to rotate, we analyze the single-disk dynamics. The side-view snapshots reveal two types of motion for an excited disk: Clattering~\cite{Goyal1998} and precession. In the clattering mode, the disk flaps [Fig.~\ref{fig:rot}(a)] with a period comparable to the vibration period. In order to maintain this mode, a projection of the disk on the vibrating plate should have a mirror symmetry along the line connecting the two consecutive colliding points. For the case of a circular disk with such a mirror symmetry, this mode is indeed favorable. Since this condition is not always given, it is much more favorable for a hexagonal disk to precess on the plane [see Fig.~\ref{fig:rot}(b)], reminiscent to an Euler's disk~\cite{Moffatt2000}. Therefore, the rotation detected from the top-view images corresponds to a projection of the precession of the disks on the horizontal plane. We note that the snapshots shown in Fig.~\ref{fig:rot} are taken without the lid and wetting liquid for a better visualization of both modes. Introducing the container lid and wetting liquid does not change the qualitative behavior.

As sketched in the inset of Fig.~\ref{fig:rot}(c), the precession of a disk is driven by the torque $\vec T=(G+2F_{\rm L})r_{\rm o}\cos\theta \,\vec e_{\rm p}$ induced by the gravity of the disk $\vec G$ and the normal force from the lid $\vec F_{\rm L}$, where $r_{\rm o}$ is the outer radius of the disk. For the ideal case without energy loss, we have $\vec T = \vec \Omega \times \vec L$ with the precession rate of the disk $\Omega$ and the angular momentum of the disk $\vec L=I_{\rm r} \Omega \sin \theta \,\vec e_{\rm r}$, where $I_{\rm r}$ is the moment of inertia of the disk about the radial direction $\vec e_{\rm r}$. Supposing the disk does not slide on the container bottom, the rotation speed $\omega$ is related to the precession rate by a factor $k=\omega/\Omega=1/\cos\theta-1$. Consequently, we have

\begin{equation}
\label{eq:omega}
\omega = k\sqrt{\frac{Gr_{\rm o}(1+2\Gamma)}{I_{\rm r}\sin \theta}},
\end{equation}

\noindent where we consider $F_{\rm L}=F_{\rm d}=\Gamma G$ for the sake of simplicity. Different from circular disks, $\Omega$ for the hexagonal disks varies discontinuously at the transition from tip to edge contacts. Moreover, the broken circular symmetry leads to a potential energy variation at fixed $\theta$, and consequently fluctuations of the rotation speed. Indeed, a close view of the disk suggests a hindered rotation during the change from edge to tip contact with the container bottom. If the rotational kinetic energy is not sufficiently large, the disk may change its sense of rotation, or switch to the clattering mode. Thus, sufficient energy injection is crucial to keep the disks excited. Otherwise, the rotating disk tends to settle down through the clattering mode, preferably with edge-to-edge contacts, because of the higher energy dissipation through inelastic collisions in this mode.

Quantitatively, we compare the angular velocity $\omega = \langle [\alpha(t)- \alpha(t-{\rm d}t)]/{\rm d}t \rangle$, where ${\rm d}t$ is the time step between subsequent frames and $\langle...\rangle$ denotes an average over all excited disks and frames captured, to the prediction of Eq.~\ref{eq:omega} in Fig.~\ref{fig:rot}(c). We assume that the disks always tend to maximize the tilting angle in the driven system. From the particle geometry and $H$, we derive $\theta = 5.8^{\circ}$ and ${\frac{I_{\rm r}}{G}=\frac{h^2}{12g}+\frac{2\sqrt{3}a^4/9-\pi d^4/64}{gA_{\rm p}}}$. Together with ${r_{\rm 0} = a/2}$, we have an analytical prediction of $\omega$ as a function of $\Gamma$, which agrees well with the experimental results without any fit parameters. This model indicates that the injected energy is more likely to be pumped into the rotational degrees of freedom as $\Gamma$ grows. Moreover, the dynamics of the disks is not controlled by the mass because $I_{\rm r} \propto G$. Instead, geometry of the disks and confinement are key parameters determining the rotational dynamics. Note that Eq.~\ref{eq:omega} applies only for excited particles with a certain tilting angle.

To summarize, we demonstrate the possibility of selectively excite the rotational degrees of freedom of individual particles. The preference to rotate arises from the broken circular symmetry of the disk shape, which leads to precession of the disks on the vibrating plate. The translational order arises from intermittent capillary interactions between neighboring disks due to the wetting liquid added. The crystalline structure is robust against variations of the agitation frequency and strength, suggesting that the detailed balance of energy injection and dissipation plays a minor role. Depending on the agitation strength and cohesion, the rotational degrees of freedom can be geometrically frustrated, giving rise to a slight compaction of the crystal and a modulation in the angle distributions. 

The dependence of rotation speed on $\Gamma$, as reveled by the analytical model, suggests the possibility to control the rotational dynamics of particles via shape design. Such a possibility sheds light on creating model systems (e.g., self-propelled `rotors') for investigating the collective dynamics of active matter~\cite{Fily2012a}. Moreover, introducing `rotors' into a granular medium can be useful in effective mixing or probing local rheology. Last but not least, the the similarity between the NESS discovered here and the RC state with translational but no orientational order triggers the question of how to define this mesophase in widespread nonequilibrium systems and how to use statistical mechanics to describe it~\cite{Sachdev1985}.

\begin{acknowledgments}
We thank Ingo Rehberg, Simeon V\"olkel, Lee Walsh, and Andreas Zippelius for helpful discussions. This work is supported by the Deutsche Forschungsgemeinschaft through Grant No.~HU1939/4-1. 
\end{acknowledgments}

%\bibliography{hex}

\end{document}